\documentclass[conference]{IEEEtran}

\usepackage[english]{babel}
\usepackage[latin1,utf8]{inputenc}
\usepackage[T1]{fontenc}
\usepackage{array}
\usepackage{amsmath}
\usepackage{amssymb}
\usepackage{bm}
\usepackage{graphicx}
\usepackage[colorinlistoftodos]{todonotes}

\title{Demo: Non-classic Interference Alignment for Downlink Cellular Networks}

\author{Yasser Fadlallah$^{1,2}$, Leonardo S. Cardoso$^{1,2}$ and Jean-Marie Gorce$^{1,2}$\\
$^{1}$University of Lyon, INRIA, France\\
$^{2}$INSA-Lyon, CITI-INRIA, F-69621, Villeurbanne, France\\
$\{$yfadlallah@gmail.com$\}$\\
$\{$jean-marie.gorce,leonardo.cardoso@insa-lyon.fr$\}$}

\begin{document}
\maketitle

\begin{abstract}
Our demo aims at proving the concept of a recent proposed interference management scheme that reduces the inter-cell interference in downlink without complex coordination, known as non -classic interference alignment (IA) scheme. We assume a case where one main Base Station (BS) needs to serve three users equipments (UE) while another BS is causing interference. The primary goal is to construct the alignment scheme ; i.e. each UE estimates the main and interfered channel coefficients, calculates the optimal interference free directions dropped by the interfering BS and feeds them back to the main BS which in turn applies a scheduling to select the best free inter-cell interference directions. Once the scheme is build, we are able to measure the total capacity of the downlink interference channel. We run the scheme in CorteXlab ; a controlled hardware facility located in Lyon, France with remotely programmable radios and multi-node processing capabilities, and we illustrate the achievable capacity gain for different channel realizations.
\end{abstract}

\section{Introduction}
Current networks aim to support high data rates to end users by increasing the spectral efficiency in bits-per-Hertz of the network, at the expense of the energy efficiency of the network. Indeed, an important part of the energy consumption of mobile networks is proportional to the radiated energy, which relies on the frequency bandwidth and the transmission power.  Any energy efficient transmission scheme should exploits the whole system bandwidth by allocating the entire available spectrum to each base station. Such an approach, however, leads to significant interference increase and performance degradation for mobiles located at the cell edges. The key challenge is to balance interference avoidance and spectrum use to reach an optimal spectral efficiency – energy efficiency (EE-SE) trade-off \cite{Tsilimantos2015}.

This challenge has been addressed in the past, for instance using frequency/code planning in 2G/3G networks or with cooperative multiple point antennas in 4G \cite{CoMP}. The aim of our work is to demonstrate the concept of a recently driven interference management scheme that not only leverage the current technology but also achieves greater overall energy efficiency \cite{GT}. It exploits Interference Alignment (IA) concept for downlink to reduce inter-cell interference without complex coordination. The theoretical achievements of IA has been largely discussed e.g. \cite{MA,CJ,AG,TG}. It has been concluded that one of the key results is that, under specific conditions, dense and high-power wireless networks are not fundamentally interference limited. As an example, under idealized assumptions, using IA in the setting of an interference channel formed by $K$ transmitter-receiver pairs interfering between one another allows an achievable data rate per pair equal to half of his interference-free channel, regardless of $K$ \cite{CJ}. Strong efforts in the  research community have been done to extend IA far beyond the initial $K$ -user interference channel. The recent review \cite{OA} highlights the different  technical challenges to be solved before envisioning a practical application  among which implementing accurate feedback loops is probably the most important  challenge. But even beyond the practical implementation of IA solutions in a network, the actual model of the network tends to be complex and involve a large number of hypothesis. These assumptions, or lack thereof, are needed and will play a  significant role in the design of IA schemes. These IA schemes are in return heavily tuned to the specific hypotheses made and may not adapt to all cellular configurations, thereby justifying the need to develop an experimental evaluation of these techniques. A downlink cellular network is basically an interfering broadcast channel, where Base Stations (BS) transmit towards a number of users and  interfere with each other. However, several tries to extend this approach  in the context of cellular networks revealed some limiting improvement \cite{DA} for the following reasons: i) The direct extension of the IC model to cellular networks relies on defining first the association of each mobile to a given subset of resources. Thus in each cell, the BS decides without coordination which set of resources is given to each mobile. To have a significant gain, IA should be performed for users mutually suffering from interference.
ii) Many works rely on a clustering approach \cite{RT}, however, users located at cluster edges cannot benefit from any improvement and are still subject to strong interference.
iii) Signaling requirements reduce the theoretical gain of the
system.

In \cite{SuhTse}, Suh et \textit{al.} proposed an IA scheme for downlink channels by considering a scenario where each BS uses a reduced space for its own transmission, preserving a given free subspace for other cells. Their solution extended in \cite{GT}, allows users in a cell to cancel their dominant interferer as well as the intra-cell interference for users in the same sub-band, achieving one degree of freedom (DoF), without any communication between the BSs. In other words, each BS create an interference-free hole in which the other BSs can serve their mobile users. Each mobile measures and feeds back its own free subspace to its main BS. After receiving this information from all associated UE the BS jointly computes a set of precoders and schedules the UE to maximize the overall capacity, under some fairness/priority constraints. Basically, a certain cooperation between cells is not explicit but exists through the feedback channels. The performance of such scheme relies on many parameters such as synchronization, feedback capabilities, precoders choice... In this paper, we review the aforementioned non-classic IA scheme for downlink proposed in \cite{SuhTse} and \cite{GT}. Then, we show the implementation scenario in the experimentation shielded room CorteXlab located in Lyon, France, and we discuss some experimental issues. Finally, we present the results of our demo and highlight the capacity gain of the implemented IA scheme over the classical OFDMA scheme.

Notations: boldface upper case letters and boldface lower case letters denote matrices and vectors, respectively. The superscripts $(.)^{T}$ and $(.)^{H}$ stand for the transpose and transpose conjugate matrices, respectively. The following notation
\begin{equation}
\bm V = \mathcal{V}_{\text{left}}\left(\text{SVD}(\bm A)\right)^{N-d+1:N},
\end{equation}
denotes the last $d$ vectors of an $N$-dimension space matrix $\bm V$ defined as the left-singular matrix of a matrix $\bm A$, where SVD($\bm A$) stands for the Singular Value Decomposition (SVD).

\section{Non-classic Interference Alignment}
\subsection{System model}
For our demo, we consider a cellular network with two BS and three user equipment (UE), all equipped with a single antenna. The transmission scheme is based on OFDM with $K$ available sub-carriers. The transmit signal of the interfering BS is written as
\begin{equation}\label{eq1}
\bm{x'}_i=\bm M_H\bm x_i,
\end{equation}
where $\bm{x'}_i$ is the data vector of length $K-N_f$ before Hadamard precoding. In the implemented scheme, each BS preserves $N_f$ DoF for the other BSs, thereby creating a hole free of interference at all cell-edges. The maximum number of transmitted streams is then $N_s=K-N_f$. $\bm M_H$ is a unitary trunk Hadamard matrix of dimension $K\times N_s$ that allows the use of the $N_s$ reduced space.

We denote $\bm H_{mb}$ the fading channel matrix between the $b^{th}$ BS and the $ m^{th}$ UE. Let the reduced channel denotes $\bm G_{mb}=\bm H_{mb}\bm M_H$, where $b=d$ and $b=i$ refer to the desired and interfering channel, respectively. The received signal at the $m^{th}$ UE is written as
\begin{equation}\label{eq2}
\bm y_m=\bm G_{md}\bm x_d+\bm G_{mi}\bm x_i+\bm w_m,
\end{equation}
where $\bm w_m$ is the noise vector. In addition to the Hadamard trunking matrix, each original stream $\bm s_{b,i}$ with $i \in \{1,\cdots ,N_s\}$ is carried over the precoding vector $\bm c_i$ ; the $i^{th}$ vector of a precoding matrix $\bm C_b$. The precoded stream vector at the $b^{th}$ BS is defined as
\begin{equation}\label{eq3}
\bm x_b=\bm C_b\bm s_b,
\end{equation}
and includes all streams associated to the $b^{th}$ BS. In order to accomplish a reliable interference management scheme, we review the non-classic IA technique extended in \cite{GT} that calculates an optimal precoding set lying over the free-interference subspace and then selects the best subset that maximizes the total capacity of the downlink interference channel.

\subsection{Non-classic IA technique}
The fading channels connecting the UEs to their main BS and to the strongest interferer BS are firstly estimated at the all UEs. This is possible through training sequences provided by both BSs. Then, each UE determines the interference null space thanks to the SVD of the reduced channel $\bm G_{mi}$. It is calculated as
\begin{equation}\label{eq4}
\bm V^\perp=\mathcal{V}_{\text{left}}\left[(\text{SVD}\left(\bm G_{mi}\right)\right)^{N_s+1:N_s+N_f}.
\end{equation}
By applying a linear projection on the interference null space $\bm V^\perp$, the received signal is rewritten as
\begin{equation}\label{eq5}
\bm V^\perp\bm y_m=\bm V^\perp\bm G_{md}\bm x_d+\bm V^\perp\bm w_m,
\end{equation}
In other words, the signal $\bm y_m$ is being received in the inter-cell interference-free hole created by the strongest interferer. This means that the best for a UE is to have its desired streams aligned with the equivalent channel $\bm G_{mb}^\perp=\bm V^\perp\bm G_{mb}$ which guarantees a maximum received power of the desired signal. It is worth noting that in practice, the UEs at cell edges can be interfered from more than one BS. Dealing with such as a case requires either considering the interference as noise or estimating all interfering channels while the transmission is focused in the joint hole created by the union of the interference. For the sake of simplicity, we only assume one interfering BS for our demo.

Each UE calculates its optimal $N_f$ precoding vectors and feeds them back to the main BS, which in turn collects $N_uN_f$ precoding vector where $N_u$ is the total number of user fixed to three in our demo. However, the maximum number of streams allowed for a BS is $N_s$. When $N_s> N_uN_f$, all vectors are used for transmission. Otherwise, i.e. $N_s<N_uN_f$, an $N_s$ precoding vectors must be selected among the total set. That is, we have to run a scheduling to select which users and streams are best to serve. Among the different existing criteria, we choose to maximize the total channel capacity given by
\begin{equation}\label{eq6}
R_d=\sum_{l=1}^{N_s}\log_2\left(1+sinr_l\right),
\end{equation}
where the term $sinr_l$ stands for the signal to interference and noise ratio at the $l^{th}$ selected stream after scheduling.

When a scheduling is required, the precoding vectors set can be presented as an underdetermined matrix of dimension $N_s \times N_uN_f$. The problem is seen as selecting the best determined matrix that maximizes the total rate. The optimal solution is through exhaustive search. For this to happen, we need to build all possible subset combination and to calculate the achievable rate of each candidate. The data symbols are carried over the vectors yielding a maximum achievable rate. The precoding vector of the $l^{th}$ stream of the $m^{th}$ user is calculated as
\begin{equation}
\bm c_l=\bm G_{md,l}^{\perp,H},
\end{equation}
where $\bm G_{md,l}$ is the $l^{th}$ row of the reduced interference-free channel. In other words, the precoding vector aims at maximizing the power of the desired part of the received signal. Hence, each user relies on the fact that its interference-free directions are independent from the interference-free directions of the other UEs associated to the same BS, which means that channel diversity between UEs has a major impact for achieving a desired performance of the described interference management scheme. That is, when the channels are almost orthogonal there is no intra-cell interference and the achievable rate is maximum. In contrast, when the channels between UEs are completely correlated, the interference reaches its maximum and the rate is highly degraded.

Going back to the scheduling problem, it can be expressed as
\begin{eqnarray}\label{eq7}
& &\arg\max R_d\left(\bm p_1, \cdots , \bm p_{N_s}\right) \nonumber \\
& &\text{s.t.}\ \ \{\bm p_1, \cdots , \bm p_{N_s}\} \in \mathcal{C}
\end{eqnarray}
where $\mathcal{C}=\{\bm c_1, \cdots , \bm c_{t}\}$ and $t=N_uN_f$. Such a problem is known as a discrete optimization that requires in most cases an exhaustive search to find out the global optimum. However, this becomes so costly and expensive with the dimension increases. Other sub-optimal algorithms based on heuristic methods or local minima search can be applied with no guarantee of any improved performance. Since we do not assume high dimensions systems and we apply non-classic IA technique over only four sub-carriers, the optimal exhaustive search is feasible with low computational cost. Once the best subset is selected, we proceed to calculate the precoding matrix of all streams that cancels the intra-cell interference. It is given by
\begin{equation}\label{eq8}
\bm V_{IA}\bm P = \bm I_{N_s},
\end{equation}
where the selected precoding vectors are the columns of the matrix $\bm P$ and $\bm I_{N_s}$ is the identity matrix. The achievable rate can be then written as
\begin{equation}\label{eq9}
R_d=\sum_{l=1}^{N_s}\log_2\left(1+\alpha_{l}sinr_l\right),
\end{equation}
where $\alpha_{l} = \frac{1}{\bm v^l}$ and $\bm v^l$ is the $l^ {th}$ column of $\bm V_{IA}$. It can be noticed that $\alpha_{l} \leq 1$ and is equal to one when the selected vectors are all orthogonal.

\begin{figure*}
\centering
\includegraphics[scale=0.3]{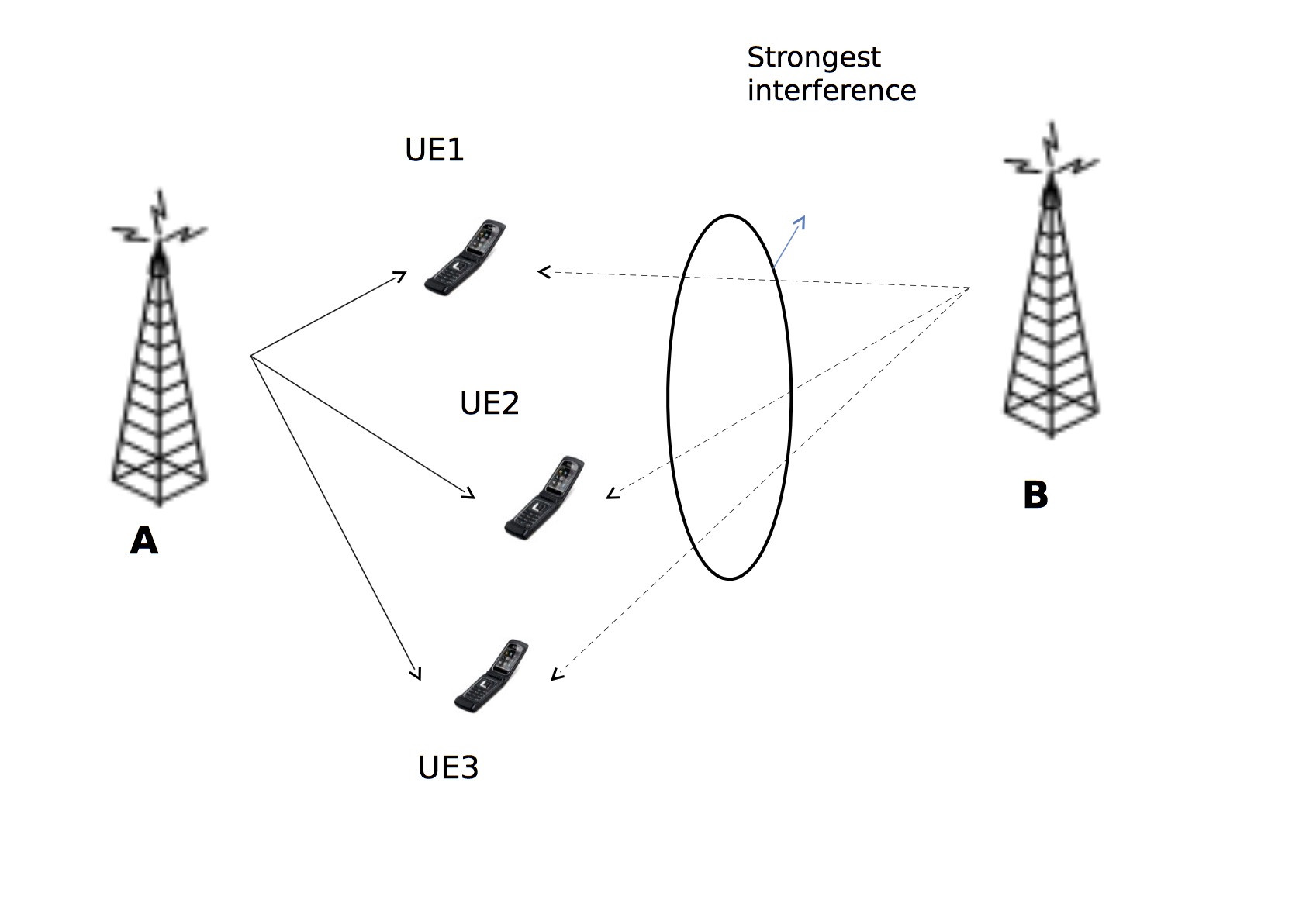}
\caption{The proposed scenario}
\label{fig1}
\end{figure*}

The performance of the addressed non-classic IA approach has been evaluated through exhaustive simulations in different networks and load scenarios. It has been found that in comparison to a reference scenario without IA, the cell capacity can on average be increased by a factor $2$ and that the spectral efficiency of cell edge users can be increased up to a factor $4$ \cite{GT}. Our work herein gives a proof of concept and focuses on the main challenges related to the non-classic IA approach for downlink, namely the knowledge of the interference footprint and the scheduling algorithms to make use of the interference information to maximize the spectral efficiency. We implement an experimental scenario with two transmitting BSs, i.e. the main BS and an interfering one and three UEs associated to the main BS. We show that the IA is feasible, in the sense that UEs are able to measure a channel, feed it back to the BS which in turn applies a scheduling to select the best set and calculates the theoretical Spectral Efficiency (SE) gain over a classical dumb OFDMA allocation.

\begin{figure*}
\centering
\includegraphics[scale=0.5]{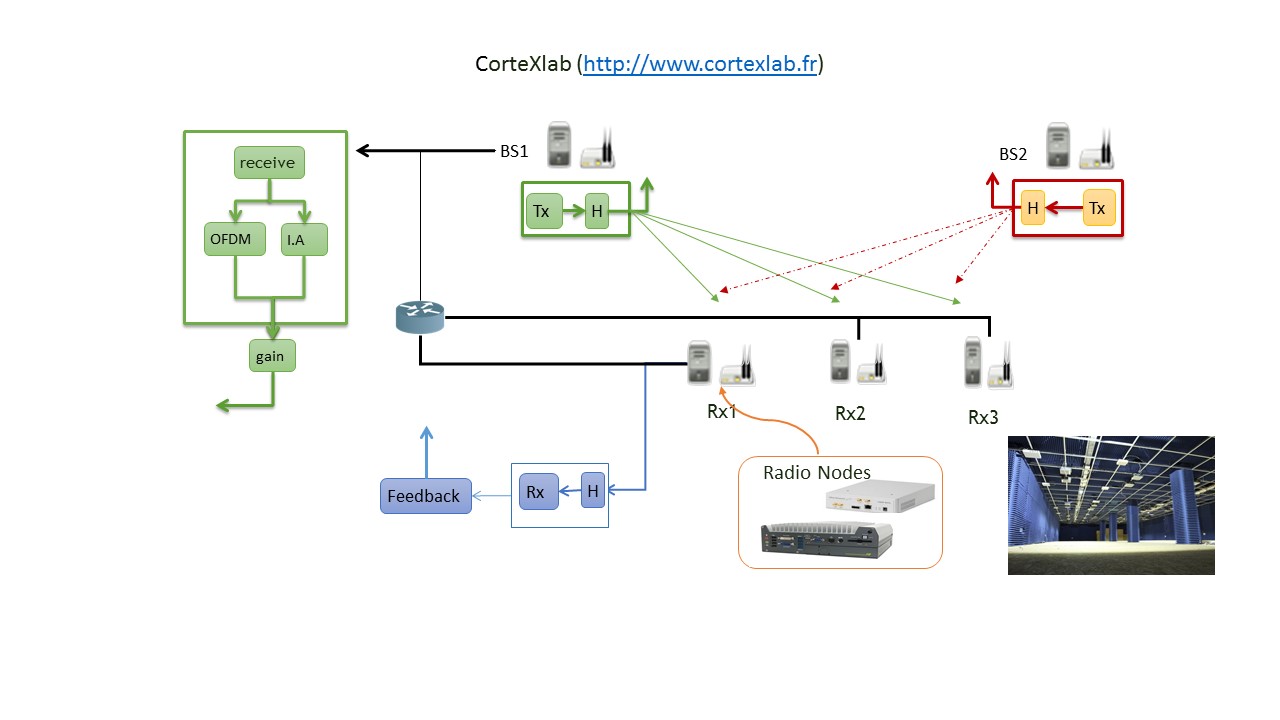}
\caption{Experimentation scheme in CorteXlab}
\label{fig2}
\end{figure*}

\section{Scenario and Experimentation}

A wireless network is emulated on CorteXlab (http://www.cortexlab.fr) \cite{cortexlab1,cortexlab2,cortexlab3}, a controlled hardware facility located in Lyon, France with remotely programmable radios and multi-node processing capabilities. During the live demo, a control laptop is remotely connected to the facility, deploying software on the radios and launching an IA scenario and collecting real-time performance feedback. The efficiency gain of IA is then shown for various experimental conditions that can be tuned from the control laptop.

\subsection{Radio Platforms}
Represented by the National Instruments USRP 2932, the general-purpose Software Defined Radio (SDR) nodes use the GNU Radio toolkit for rapid prototyping of transmission techniques mostly reliant on the general purpose processor (GPP) of the host PC. The USRP 2932 is a high end radio platform, counting with a 400 MHz – 4.4 GHz RF board, data rates of up to 40 MHz (with reduced dynamic range, nominal band of 20 MHz), a precise OCXO clock source and a $1$ gigabit ethernet (GigE) connection to the host PC. The host PC is based on a Linux environment and allow users to test PHY and MAC layer techniques. It is preferable to first use both Linux and GNU Radio for the development and test at the user’s own computer and then to bring the experiment over to CorteXlab.

\subsection{Implemented scenario}
Among the forty available SDR nodes in CorteXlab, five are selected where two serve as main and interfering BS and the remainders serve as mobile UEs. All transmitters (TX) and receivers (RX) are equipped with single antenna. Therefore, the IA scheme lies on the frequency dimensions provided by the OFDM transmission scheme implemented at all TXs and RXs. In the first transmission phase, the UEs need to estimate the main and interfering channels. This can be done by assigning two orthogonal time slots to the BSs through which they transmit the training sequences. However, the synchronization of the two BS nodes remains a challenge since they are distantly separated. Herein, we propose an over-air synchronization as follows. Holding a unique ID, the interfering BS starts the transmission to let the user nodes measure the interference channel coefficients while the main BS tries to decode the ID of the transmitter. If the decoded ID corresponds to the interferer, the main BS node transmit OFDM symbols (one or more) for channel estimation. By collecting both estimated channels, each receiver is able to calculate the $d$ optimal precoding vectors in the free dimensions such that the desired signal power is maximized. In order to avoid imperfections in the IA scheme, we use a wire connection provided in the experimentation room between all nodes to perform a perfect feedback to the TX (main BS). This latter gathers the precoding vectors from all UEs and run the scheduling algorithm to seek the best $(N-N)$ precoding vectors that minimizes the intra-cell interference, and hence maximizes the achievable data rate of the cell. The maximized rate is then compared to the classical OFDMA and the theoretical gain is calculated. The OFDMA scheduler select at each users the $L_u$ streams that maximizes the Signal-to-Interference and Noise Ratio (SINR) given by
\begin{equation}\label{eq10}
\rho_{u,l} = \frac{|h_{u,l}|^2}{|h_{i,l}|^2+\sigma_n^2},
\end{equation}
where $\rho$ stands for the SINR, $h_{u,l}$ is the channel coefficient between the main BS and the $u^{th}$ user at the $l^{th}$ stream, $h_{i,l}$ is the interference channel coefficient at the $l^{th}$ stream and $\sigma_n^2$ is the noise variance. The classical OFDMA achievable rate is then defined as
\begin{equation}\label{eq11}
R_{ref} = \sum_{u=1}^{N_u} \sum_{l=1}^{L_u} \log_2\left(1+\rho_{u,l}\right).
\end{equation}
\subsection{Experimentation and results}
\begin{figure}
\centering
\includegraphics[scale=0.2]{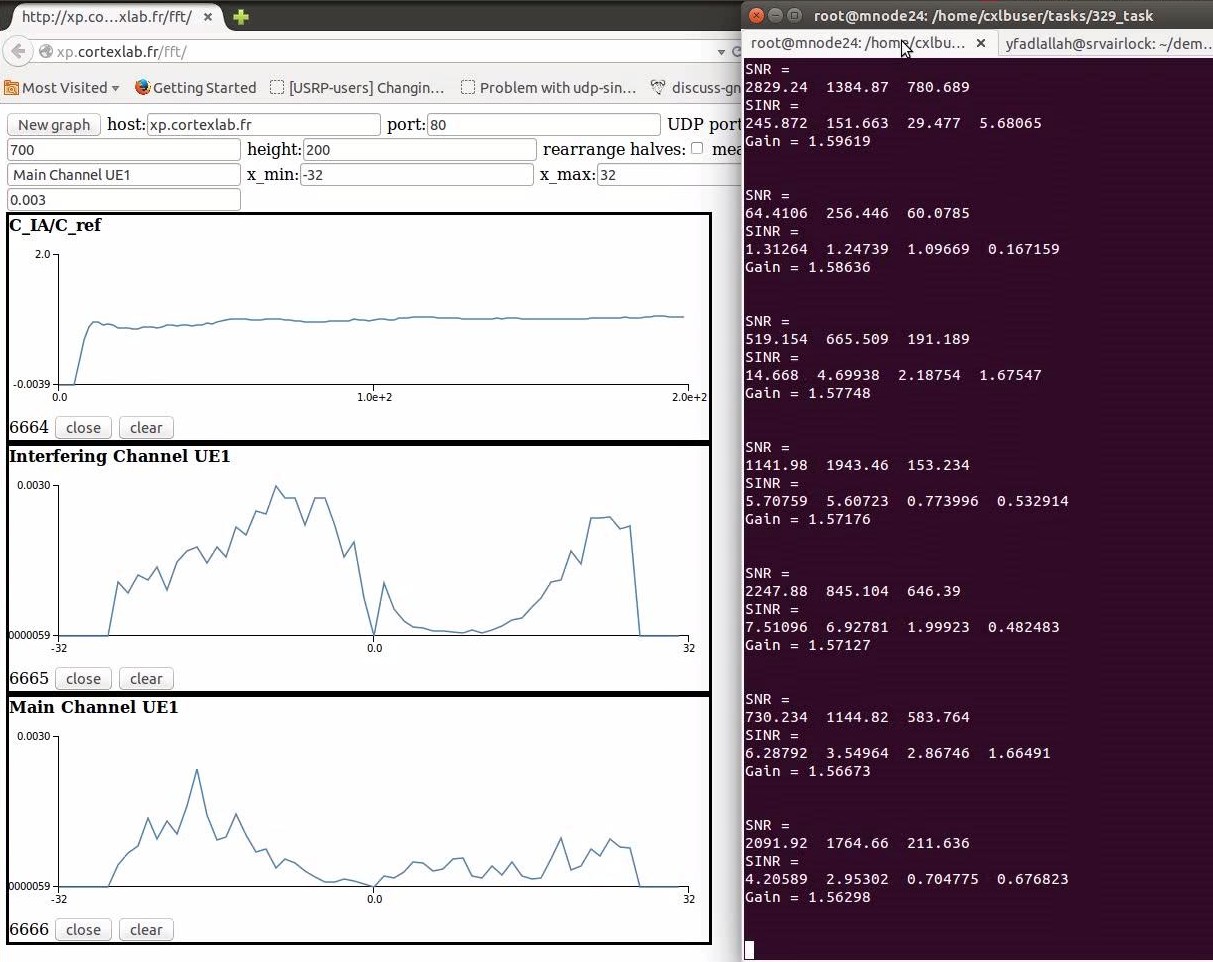}
\caption{Experimentation results}
\label{fig3}
\end{figure}

In the live demo, we show a spectral efficiency gain that can largely vary between $1$ and $3$. This variation is due to the influence of many factors summarized by the following parameters: the noise power, the interference power, the channel diversity, the distance between the different nodes, the TX gain... For instance, in perfect conditions the ratio gain $R_d/R_{ref}$ given in (\ref{eq9}) and (\ref{eq11}), tends to its maximum when the inter-cell interference power is of the same order or higher than the desired signal power ; i.e. SINR is low, this is the case of cell edge mobile users. However, as long as the users get closer to the main BS, this ratio decreases. Another critical parameter that highly impacts the performance gain is the channel diversity and correlation, the less the channel is correlated the higher the efficiency gain is. This is because the applied IA scheme requires a completely decorrelated channel coefficients for the scheduling, otherwise the intra-cell interference dominates the desired signal. In order to get a better decorrelation in the shielded room, we emulate a virtual channel on all TXs and RXs, however, this still induces some correlation at the different mobile UEs. A better way is to use a multi-antennas node at the TX to generate multi-paths through the different antennas. Each path is randomly attenuated, phase shifted and delayed. Such a multi-paths generation creates a perfect decorrelation since the paths generated by the different nodes are totally independent. An illustration of our demo is given in Figure \ref{fig2}. In our demo, we assume two BSs and $N_u=3$ users. The IA is applied over four sub-carriers which gives a total dimension of $4$, one is free of interference and three other used by each BS. We focus on the average theoretical capacity gain offered by the non-classical IA over the classical OFDMA scheme with respect to the number of transmission as shown in Figure \ref{fig3}. We also plot the channel spectrum at the all sub-carriers and for the different transmission to show how decorrelated the channel coefficients are. For each channel realization, we display the SNR of the interference-free streams and the SINR of all streams when the classical OFDMA is applied to see their impact over the efficiency gain.

\section{Conclusion and Work under-development}
In this demo, we have implemented the first phase of a non-classic IA approach for interference management. It consists in measuring the channel coefficients, calculating the interference-free sub-spaces, feeding them back to the main BS, and applying a scheduling to decide which stream to transmit. We have shown in experiment that a significant theoretical capacity gain over the classical OFDMA scheme can be achieved depending on the channel conditions and the interference dominance. The remaining work is to start the IA transmission and to apply linear decoding criteria at the receiver such as Zero-Forcing (ZF) or Minimum Mean Squared Error (MMSE) to recover the original data. However, the challenge here is to synchronize the transmission between both TXs for the IA transmission and to compare the theoretical gain to the practical one. It is also worth trying to face more practical issues related to de-synchronization, which means to study the impact of the delay and phase shift between the different TXs on the IA scheme.

\end{document}